# Spread of SARS-CoV-2 in a SIS model with vaccination and breakthrough infection

Ariel Félix Gualtieri[*], Carolina de la Cal, Augusto Francisco Toma, Pedro Hecht

*Universidad de Buenos Aires. Facultad de Odontología. Cátedra de Biofísica y Bioestadística. Buenos Aires, Argentina.*

## Abstract

Although previous infection and vaccination provide protection against SARS-CoV-2 infection, both reinfection and breakthrough infection are possible events whose occurrence would increase with time after first exposure to the antigen and with the emergence of new variants of the virus. Periodic vaccination could counteract this decline in protection. In the present work, our aim was to develop and explore a model of SARS-CoV-2 spread with vaccination, reinfection and breakthrough infection. A modified deterministic SIS (Susceptible-Infected-Susceptible) model represented by a system of differential equations was designed. As in any SIS model, the population was divided into susceptible and infected individuals. But in our design, susceptible individuals were, in turn, grouped into three consecutive categories whose susceptibility increases with time after infection or vaccination. The model was studied by means of computer simulations, which were analysed qualitatively. The results obtained show that the prevalence, after oscillating between peaks and valleys, reaches a plateau phase. Moreover, as might be expected, the magnitude of the peaks and plateaus increases as the infection rate rises, the vaccination rate decreases and the rate of decay of protection conferred by vaccination or previous infection increases. Therefore, the present study suggests that, at least under certain conditions, the spread of SARS-CoV-2, although it could experience fluctuations, would finally evolve into an endemic form, with a more or less stable prevalence that would depend on the levels of infection and vaccination, and on the kinetics of post-infection and post-vaccination protection. However, it should be kept in mind that our development is a theoretical scheme with many limitations. For this reason, its predictions should be considered with great care.

---

[*]Author for correspondence. Universidad de Buenos Aires. Facultad de Odontología. Cátedra de Biofísica y Bioestadística. Marcelo T. de Alvear 2142. Piso 17 B. (C1122AAH) Ciudad Autónoma de Buenos Aires, Argentina. Email: ariel.gualtieri@odontologia.uba.ar

---

## 1. Introduction

Since the first case of coronavirus disease-19 (COVID-19) was identified in December 2019 until the end of May 2022, the severe acute respiratory syndrome coronavirus 2 (SARS-CoV-2) infected about 530 million people and caused the death of about 6.3 million [1]. But, on the other hand, there are about 500 million recovered people [1] and 3 800 million vaccinated [2].

Previous infection and vaccination confer protection against SARS-CoV-2 infection [3,4]. However, both reinfection and infection of vaccinated individuals (breakthrough infection) may occur [3,5,6]. Antibodies generated by previous infection, as well as those produced by vaccination, wane over time [7,8]. Therefore, the risk of re-infection or breakthrough infection would increase as time passes [9]. In turn, the emergence of new virus variants would also contribute to raise the likelihood of reinfection and to reduce the effectiveness of vaccines [10-12].

Booster vaccination could raise neutralizing antibody levels [13-16], increase vaccination effectiveness [17] and reduce the magnitude of outbreaks [18]. Thus, depending on the epidemiological scenario, periodic inoculations, including vaccines against new variants, might be necessary [19-21].

Epidemic models are formal designs that capture the general behaviour of the spread of infectious diseases [22,23]. They can thus assess the influence of different variables on epidemiological dynamics. In particular, models could tentatively predict the outcomes of health intervention strategies [24]. In an almost premonitory review on models and global spread of diseases, published about a year before the first case of COVID-19 was reported, Walters *et al*. (2018) already discussed the scope and limitations of using models in the response to potential pandemics [25].

The design of an epidemic model starts by dividing the population into compartments according to different disease stages or other relevant variables. The transition dynamics between the different compartments are then represented [26]. For example, in a SIR model the population is divided into three compartments: susceptible (S), infected (I) and recovered with immunity (R). According to this grouping, the transitions S→I (infection) and I→R (recovery) are mathematically represented. In a SIS model (Susceptible→Infected→Susceptible), on the other hand, there is no permanent immunity: after individuals overcome the infection, they become susceptible again [27].

The basic reproduction number ($R_0$) is an important measure for predicting the dynamics of the spread of an infectious disease, and it is essential for the development of epidemic models. $R_0$ is defined as the average number of new infections that are caused by an infected individual within a population where all other individuals are susceptible [28]. Thus, when $R_0$ is greater than 1, the outbreak spreads. In contrast, if $R_0$ is less than 1, the epidemic does not progress [29].

The $R_0$ of SARS-CoV-2 would be quite variable. In a meta-analysis, Alimohamadi *et al*. (2020) found a range of $R_0$ from 1.9 to 6.49, and estimated a pooled $R_0$ of 3.32, with a 95% confidence interval (95% CI) of 2.81 to 3.82 [30]. In another meta-analysis, based on data collected in August 2020, Yu *et al*. (2021) estimated an overall $R_0$ of 4.08 (95% CI, 3.09-5.39) [31]. Different variants may have different $R_0$ values. In a review on $R_0$ of Delta variant (B.1.617.2), Liu & Rocklöv (2021) found a range from 3.2 to 8, with a mean of 5.08 [32]. Meanwhile, $R_0$ of Omicron variant (B.1.1.529) would be, on average, 2.5 times higher than that of Delta variant [33].

Since the start of the pandemic, numerous COVID-19 epidemic models have been devised, which have been based on different schemes [34-36], including the SIS scheme [37-39].

The objective of the present work has been to design and explore a SIS model of SARS-CoV-2 spread, with reinfection, vaccination and breakthrough infection.

## 2. Materials and methods

### 2.1. The model

A deterministic model was developed to simulate the spread of SARS-CoV-2. The model is represented by the following system of 4 differential equations:

$$\frac{dS_L}{dt} = -\frac{\beta_L S_L I}{N} + \gamma I + \nu S_H - \tau_1 S_L$$

$$\frac{dS_M}{dt} = -\frac{\beta_M S_M I}{N} + \tau_1 S_L - \tau_2 S_M$$

$$\frac{dS_H}{dt} = -\frac{\beta_H S_H I}{N} + \tau_2 S_M - \nu S_H$$

$$\frac{dI}{dt} = \frac{(\beta_L S_L + \beta_M S_M + \beta_H S_H)I}{N} - \gamma I$$

A modified SIS scheme was considered. Hence, the population was partitioned into susceptible ($S$) and infected ($I$) individuals. In turn, susceptible people were divided into three consecutive categories with an increasing degree of susceptibility. The greater the susceptibility, the greater the likelihood of getting infected. Arbitrarily, the three degrees of increasing susceptibility were named low ($L$), moderate ($M$) and high ($H$).

Thus, $S_L(t)$, $S_M(t)$ and $S_H(t)$ represent the number of susceptible individuals with low, moderate and high susceptibilities at time $t$, respectively. Similarly, $I(t)$ denotes the number of infected people. $N(t)$ is the total number of individuals in the population:

$$N(t) = S_L(t) + S_M(t) + S_H(t) + I(t)$$

Therefore, the proportion of infected individuals in the population at a time $t$, $i(t)$, is given by the following ratio:

$$i(t) = I(t)/N(t)$$

The *per capita* infection rate or force of infection, $\lambda$, is assumed to depend on the proportion of infected individuals: $\lambda = \beta I/N$, where $\beta$ is an infection parameter proportional to the contact rate and the probability of transmission [22].

According to the degree of susceptibility, the three categories of susceptible individuals are ordered as follows: $S_L < S_M < S_H$. In terms of the model, this means that the infection parameter $\beta$ increases from $S_L$ to $S_H$: $\beta_L < \beta_M < \beta_H$.

$S_H$ individuals are vaccinated with a given vaccination rate $v$. $I$ individuals recover with a recovery rate $\gamma$. Following vaccination or recovery, it is assumed that people will have a low susceptibility, meaning that they will become $S_L$ individuals. After a certain period $p_1$, however, it is supposed that the protection provided by previous infection or vaccination will wane, and that the susceptibility of the people will therefore increase to a moderate level. In terms of the model, this means that $S_L$ individuals become $S_M$ ones with a certain rate $\tau_1$ such that $\tau_1 = 1/p_1$. Likewise, it is assumed that after a certain period $p_2$, the susceptibility of $S_M$ individuals will increase to a high level. Thus, $S_M$ individuals become $S_H$ with a rate $\tau_2$ such that $\tau_2 = 1/p_2$.

The diagram of the model is shown in Figure 1.

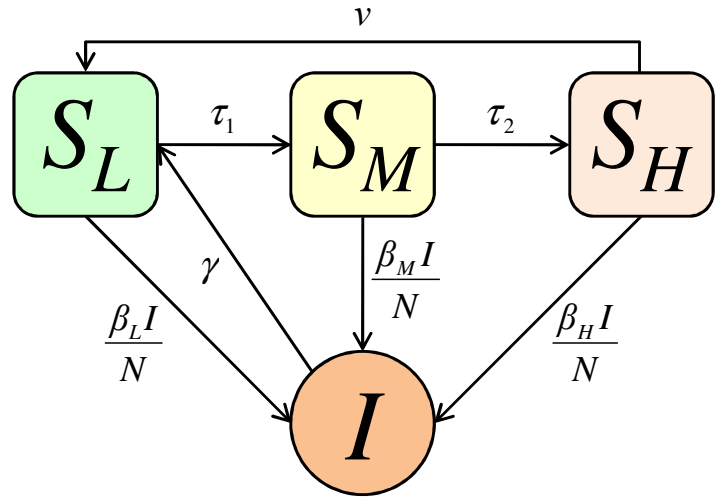

**Figure 1. Schematic flow chart of the model.**

## 2.2. Benchmark case

A benchmark case was defined. Its parameters were estimated from the range of values found in the literature, as detailed below.

An $R_0$ of 4 was chosen [30,31,33]. It was assumed $R_0 = \beta/\gamma$ [22]. An infectious period of 14 days was adopted [40]. The recovery rate, $\gamma$, is the inverse of the infectious period [41], so a value of 0.07 1/day was used for $\gamma$. Thus, for $R_0 = 4$ and $\gamma = 0.07$ 1/day, it follows that $\beta = 0.28$ 1/day. It was supposed that after either previous infection or vaccination, 90% protection against infection is achieved [3,10,42-46]. Therefore, it results that $\beta_L$ is 10% of $\beta$. That is, $\beta_L = 0.028$ 1/day.

It was presumed that 90 and 180 days after either previous infection or vaccination, protection against infection is 75% and 50%, respectively [7,10,47-49]. These assumptions lead to $p_1 = 90$ days, $p_2 = 90$ days and to the following values (1/day): $\tau_1 = 0.011$, $\tau_2 = 0.011$, $\beta_M = 0.070$ and $\beta_H = 0.140$. A vaccination rate $v$ of 0.004 1/day was used. This would be within the range of the vaccination rate in South America between January and March 2022, which was derived from the number of daily doses administered during that period and the population size according to the website *covidvax.live* [2]. We assumed that the value of $v$ can be estimated as the ratio between the number of daily doses ($\delta$) and the population size ($\psi$): $v = \delta/\psi$. The population size of South America was considered to be 433 953 687 [2].

Arbitrarily, the following initial conditions were set: $S_L(0) = 0$, $S_M(0) = 0$, $S_H(0) = 990000$, $I(0) = 10000$.

## 2.3. Numerical simulations

The dynamics of the model was explored through numerical simulations. The effects of $\beta_L$, $\beta_M$, $\beta_H$, $v$, $\tau_1$ and $\tau_2$ on the temporal evolution of the prevalence in the population, $i(t)$, were studied. When one or more parameters were changed, the rest of the conditions remained the same as in the benchmark case. The results were analyzed qualitatively. Simulations and plots were performed using R software, version 4.1.1 [50]. The R *deSolve* package was used for the simulations [51]. The plots were performed with R base packages and R *ggplot2* package [52].

# 3. Results

Figure 2 shows the evolution of prevalence as a function of time within the benchmark case. At the beginning of the simulation, the proportion of infected people, $i$, is 0.01. From the beginning of the simulation, $i$ increases until it reaches a peak of about 0.14 at roughly 70 days. After that, $i$ starts to decrease and reaches a minimum of 0.04 at about 230 days. Then $i$ increases again and, at around 400 days, reaches a lower peak than the previous one of about 0.055. Finally, $i$ decreases again and seems to stabilise roughly at 0.052.

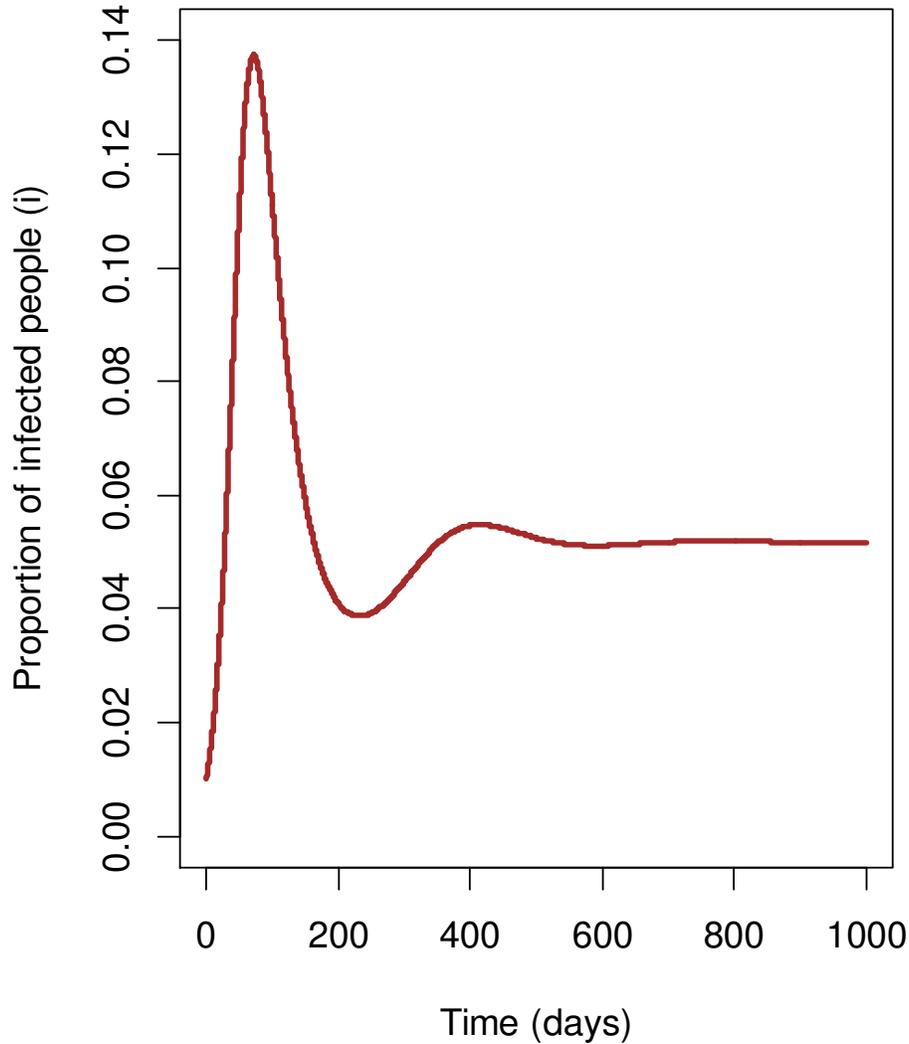

**Figure 2. Evolution of prevalence in the benchmark case. Initial conditions:** $S_L(0)=0$, $S_M(0)=0$, $S_H(0)=990\,000$, $I(0)=10\,000$. **Parameters (1/day):** $\beta_L=0.028$, $\beta_M=0.070$, $\beta_H=0.140$, $\gamma=0.07$, $\nu=0.004$, $\tau_1=0.011$, $\tau_2=0.011$.

Figures 3 and 4 allow to relate, within the benchmark case, the evolution of infected people to the dynamics of the other categories. Figure 3 shows the number of people as a function of time, while figure 4 shows the corresponding proportions.

The temporal behaviour of the four categories oscillates but tends to stabilise. During a first stage of the simulation—from the beginning until around 70 days—while $S_L$, $S_M$ and $I$ increase, $S_H$ decreases. The reduction in $S_H$ is explained by vaccination and infection. Thus, around 140 days, $S_H$ reaches a minimum of about 190000 individuals (proportion $\approx 0.20$).

Vaccination and the increase in $I$, in turn, lead to an increase in $S_L$. The increase in $S_L$ produces a consequent increase in $S_M$. Thus, after the peak of $I$ (number $\approx$ 140000; proportion $\approx$ 0.14; day $\approx$ 70) there is a peak of $S_L$ (number $\approx$ 500000; proportion $\approx$ 0.50; day $\approx$ 120) and subsequently a peak of $S_M$ (number $\approx$ 300000; proportion $\approx$ 0.30; day $\approx$ 220).

After reaching the minimum, $S_H$ increases. Meanwhile, after reaching their main peaks, $S_L$, $S_M$ and $I$ decrease. Finally, after a weak fluctuation, the dynamics of the four compartments more or less stabilises with around 380000 (38%), 290000 (29%), 280000 (28%) and 50000 (5%) individuals in the stages $S_L$, $S_M$, $S_H$ and $I$, respectively.

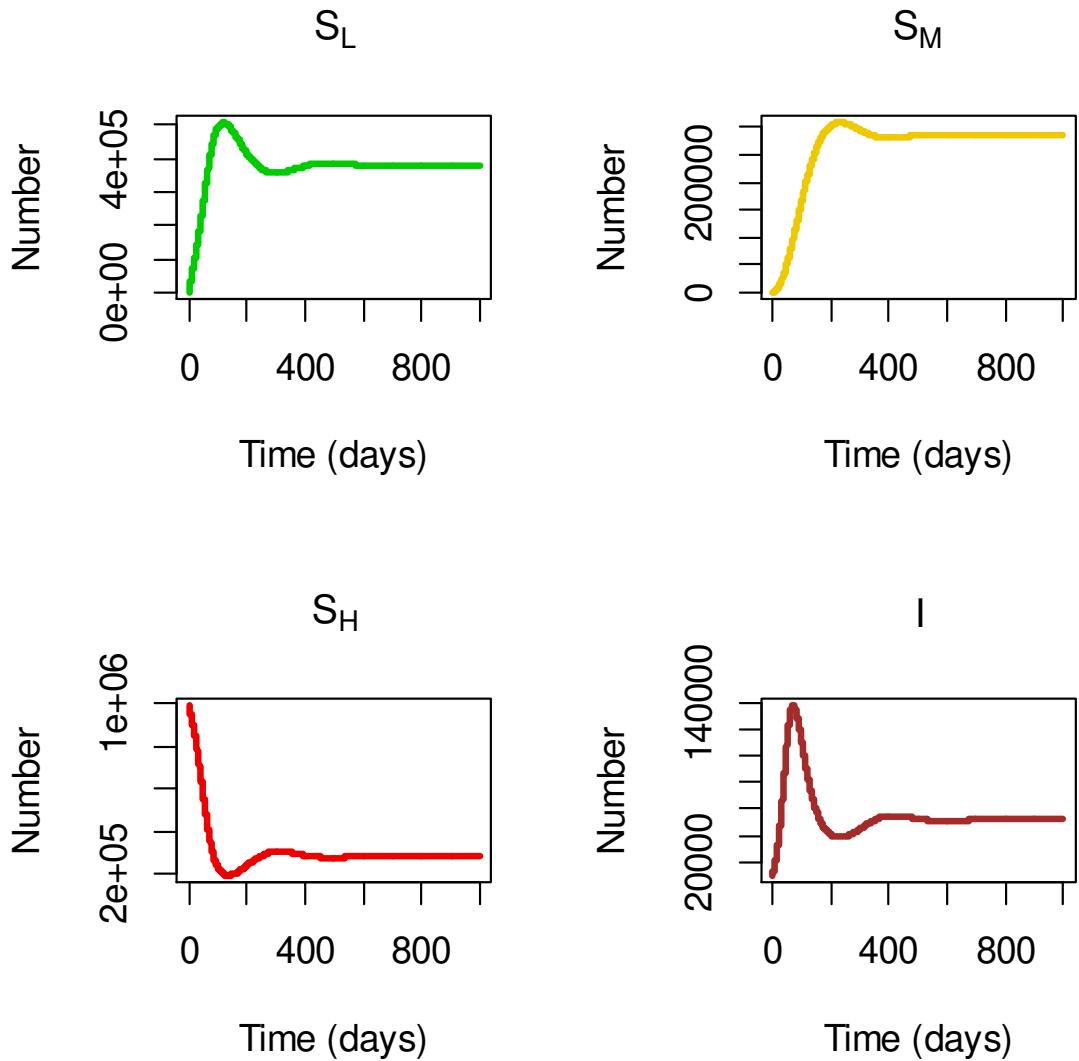

**Figure 3. Evolution of the number of individuals in stages $S_L$, $S_M$, $S_H$ and $I$ in the benchmark case. Initial conditions: $S_L(0)$=0, $S_M(0)$=0, $S_H(0)$=990000, $I(0)$=10000. Parameters (1/day): $\beta_L$=0.028, $\beta_M$=0.070, $\beta_H$=0.140, $\gamma$=0.07, $\nu$=0.004, $\tau_1$=0.011, $\tau_2$=0.011.**

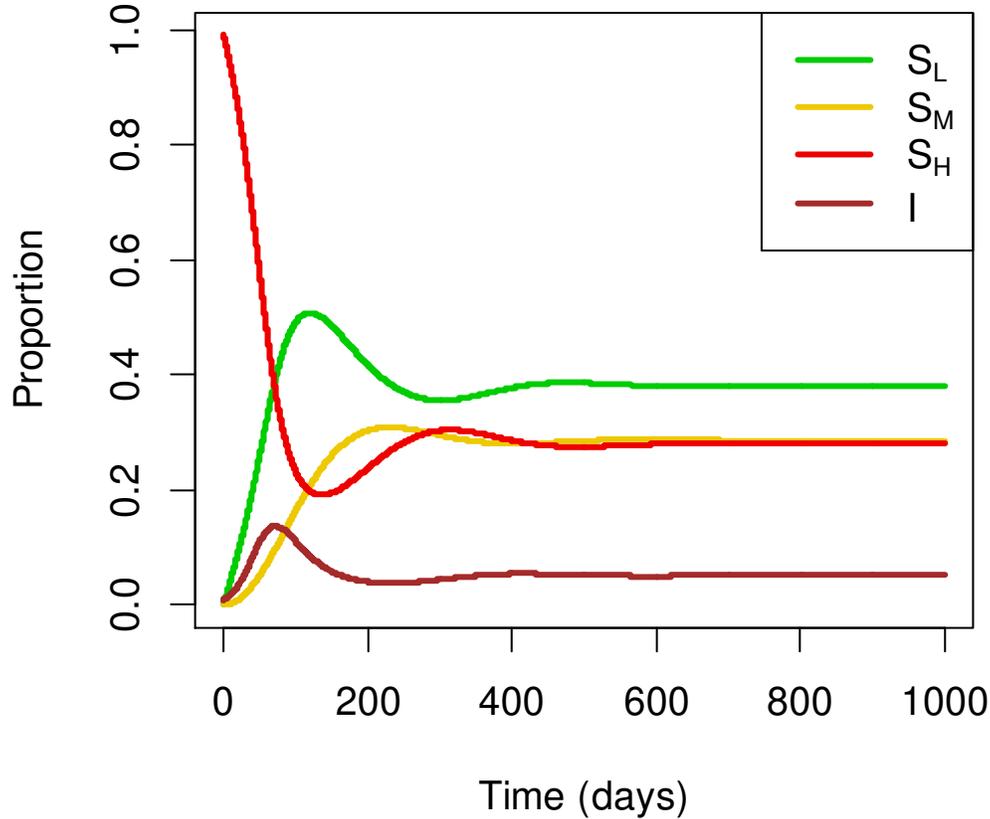

**Figure 4. Evolution of the proportion of individuals in stages $S_L$, $S_M$, $S_H$ and $I$ in the benchmark case. Initial conditions: $S_L(0)=0$, $S_M(0)=0$, $S_H(0)=990000$, $I(0)=10000$. Parameters (1/day): $\beta_L=0.028$, $\beta_M=0.070$, $\beta_H=0.140$, $\gamma=0.07$, $v=0.004$, $\tau_1=0.011$, $\tau_2=0.011$.**

The influence of $\beta_L$, $\beta_M$, $\beta_H$, $v$, $\tau_1$ and $\tau_2$ on the proportion of infected people, $i$, is depicted in figures 5 to 12. The independent increases in $\beta_L$, $\beta_M$ or $\beta_H$ causes an overall rise in the prevalence curve: the two peaks increase, the minimum in between them grows and a higher plateau occurs (Figures 5 to 7). Increasing all three parameters together produces the same effect (Figure 8).

Reducing the vaccination rate causes an increase in both peaks of prevalence and also leads to a plateau with more infected people (Figure 9). In the particular case of no vaccination ($v=0$) the first peak of prevalence reaches a value of about 0.20, which represents an increase of about 40% compared to the reference case.

Separate increases in $\tau_1$ or $\tau_2$ cause an overall increase in prevalence along the entire curve and an earlier second peak (Figures 10 and 11). Increasing $\tau_1$ and $\tau_2$ together produces the same response (Figure 12). As mentioned in the section of materials and methods, $\tau_1$ and $\tau_2$ are the inverse of $p_1$ and $p_2$, respectively. Thus, an increase in $\tau_1$ and $\tau_2$ corresponds to a decrease in $p_1$ and $p_2$, respectively.

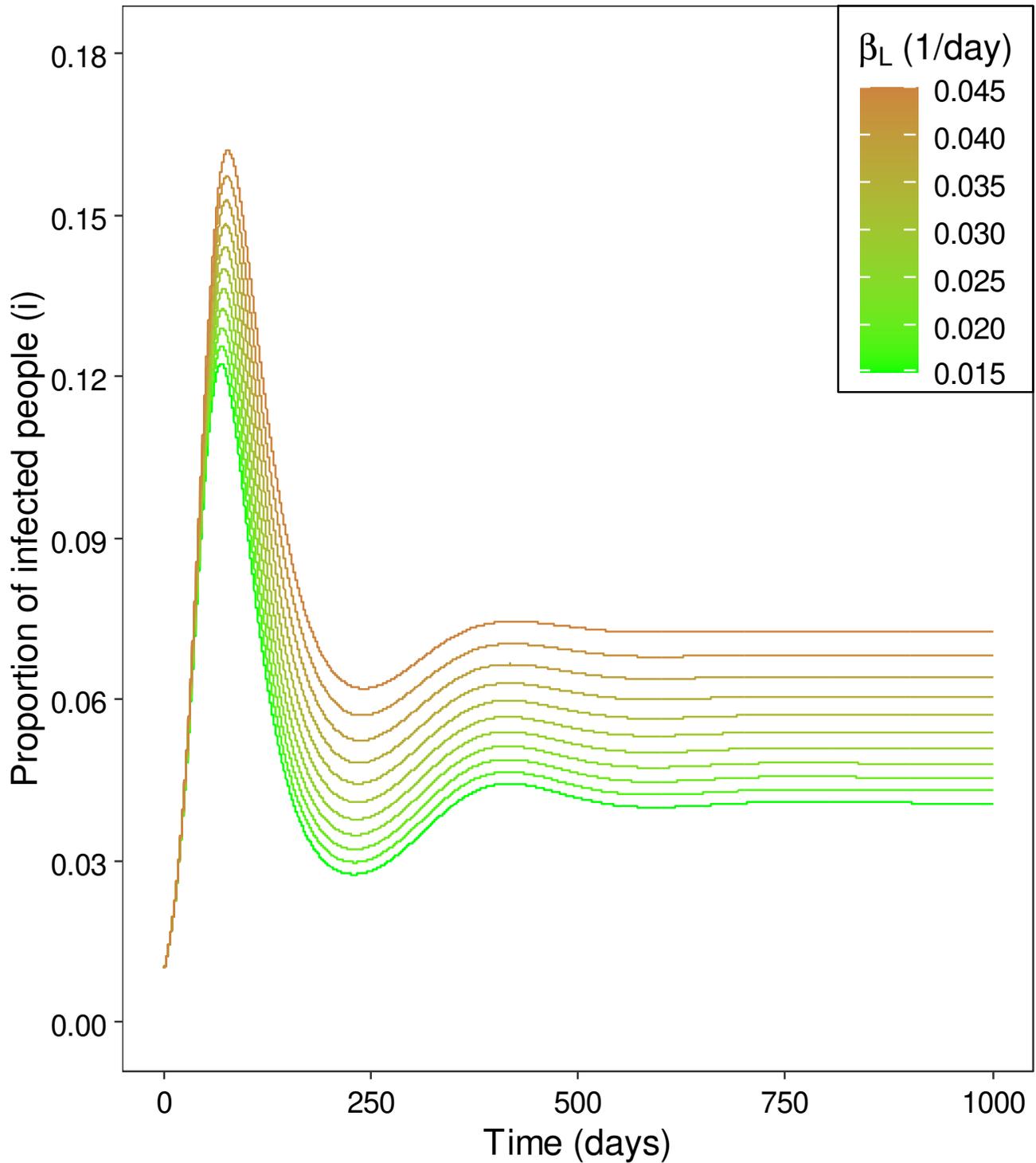

**Figure 5. Influence of the infection parameter associated to stage $S_L$ ($\beta_L$) on the evolution of the prevalence ($i$). $\beta_L$ (1/day)={0.015 to 0.045, by increments of 0.003}. The rest of the conditions were maintained as in the benchmark case.**

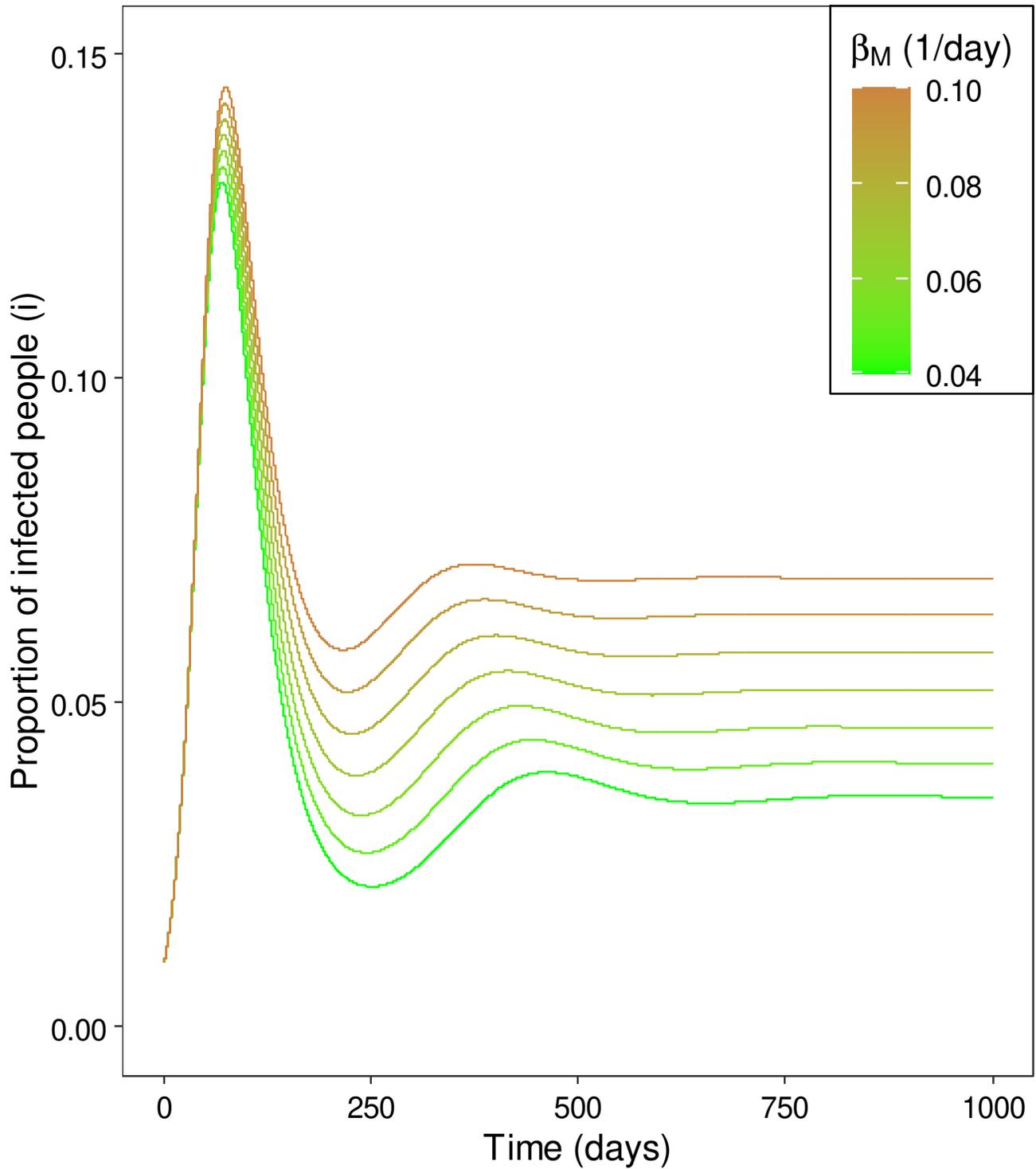

**Figure 6.** Influence of the infection parameter associated to stage $S_M$ ($\beta_M$) on the evolution of the prevalence (*i*). $\beta_M$ (1/day)={0.04 to 0.10, by increments of 0.01}. The rest of the conditions were maintained as in the benchmark case.

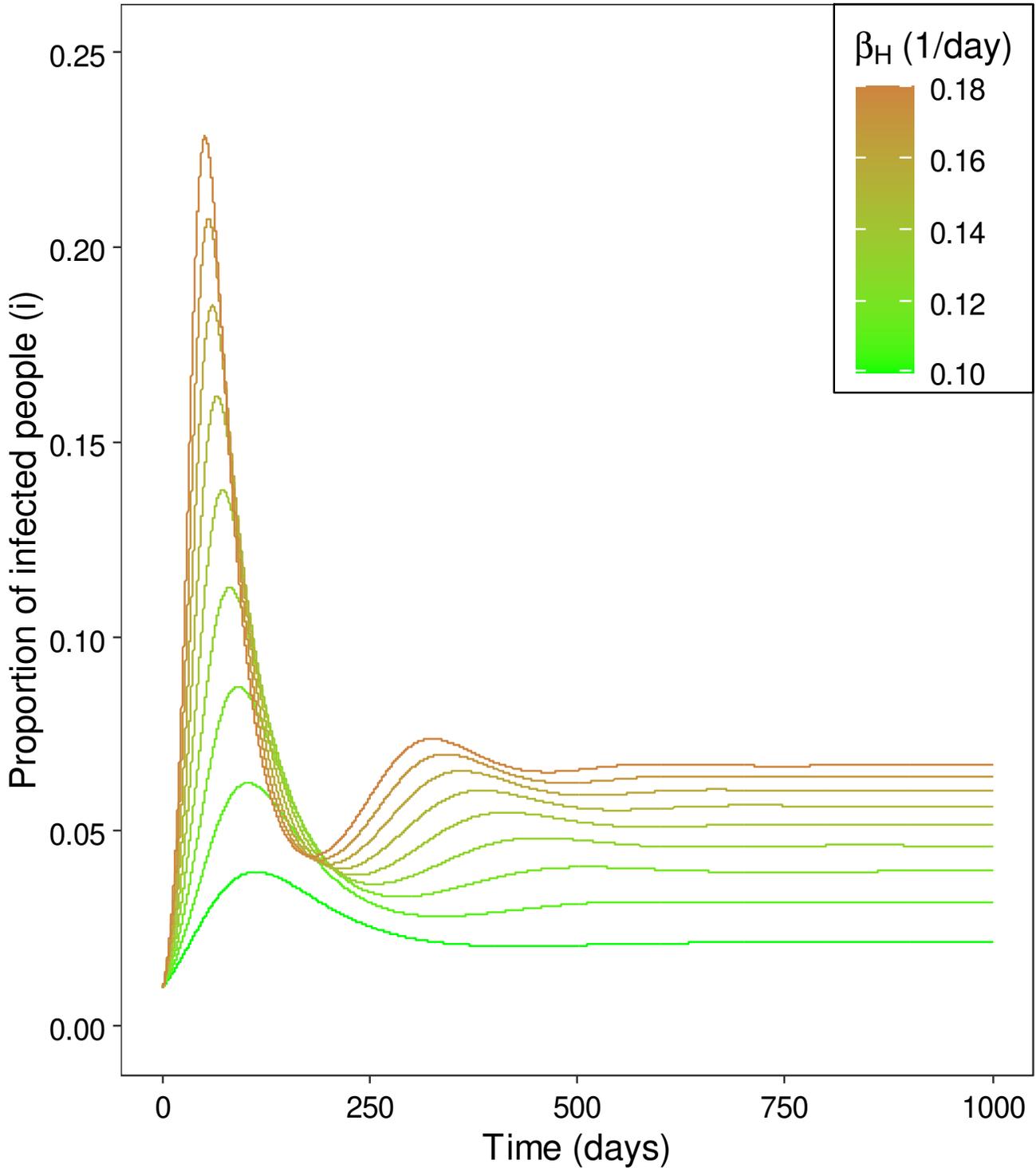

**Figure 7. Influence of the infection parameter associated to stage** $S_H$ **(**$\beta_H$**) on the evolution of the prevalence (***i***).** $\beta_H$ **(1/day)={0.10 to 0.18, by increments of 0.01}. The rest of the conditions were maintained as in the benchmark case.**

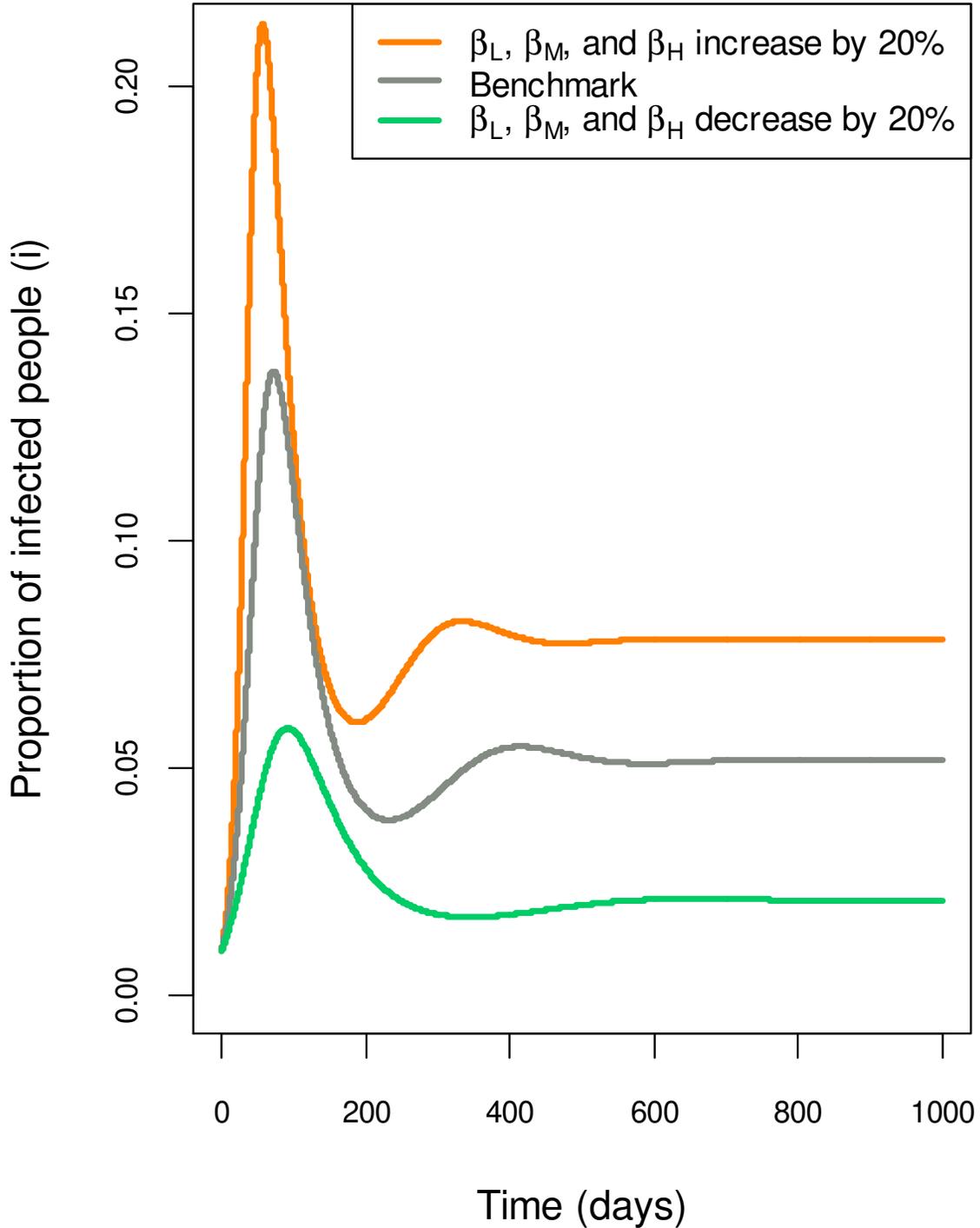

**Figure 8. Aggregate influence of the three infection parameters ($\beta_L$, $\beta_M$ and $\beta_H$, 1/day) on the evolution of prevalence (*i*). Grey, benchmark: $\beta_L$=0.028, $\beta_M$=0.070, $\beta_H$=0.140. Red, 20% increase in the three infection parameters: $\beta_L$=0.034, $\beta_M$=0.084, $\beta_H$=0.168. Green, 20% decrease in the three infection parameters: $\beta_L$=0.022, $\beta_M$=0.056, $\beta_H$=0.112. The rest of the conditions were maintained as in the benchmark case.**

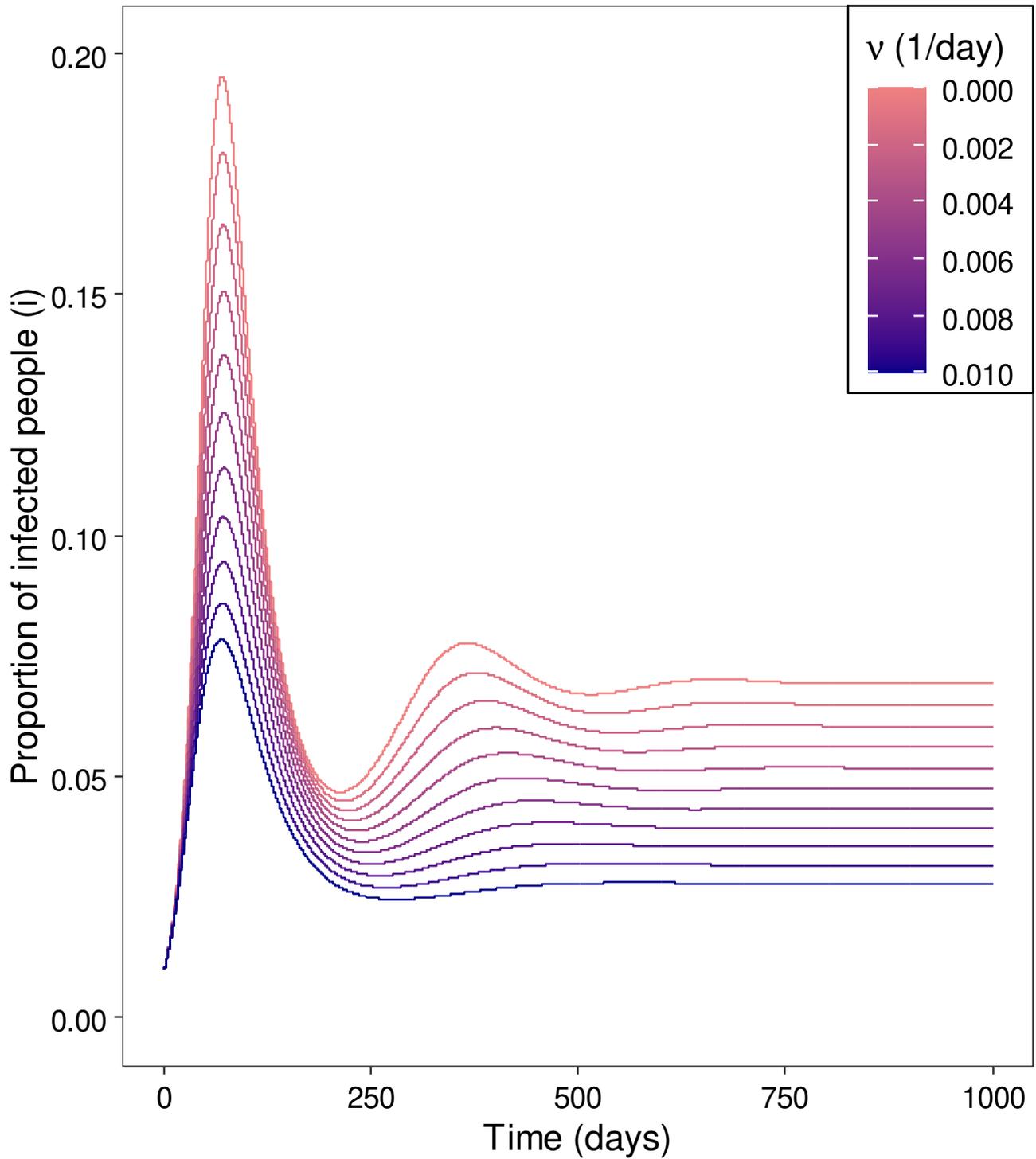

**Figure 9. Influence of the vaccination rate (*v*) on the evolution of the prevalence (*i*). *v* (1/day)={0.000 to 0.010, by increments of 0.001}. The rest of the conditions were maintained as in the benchmark case.**

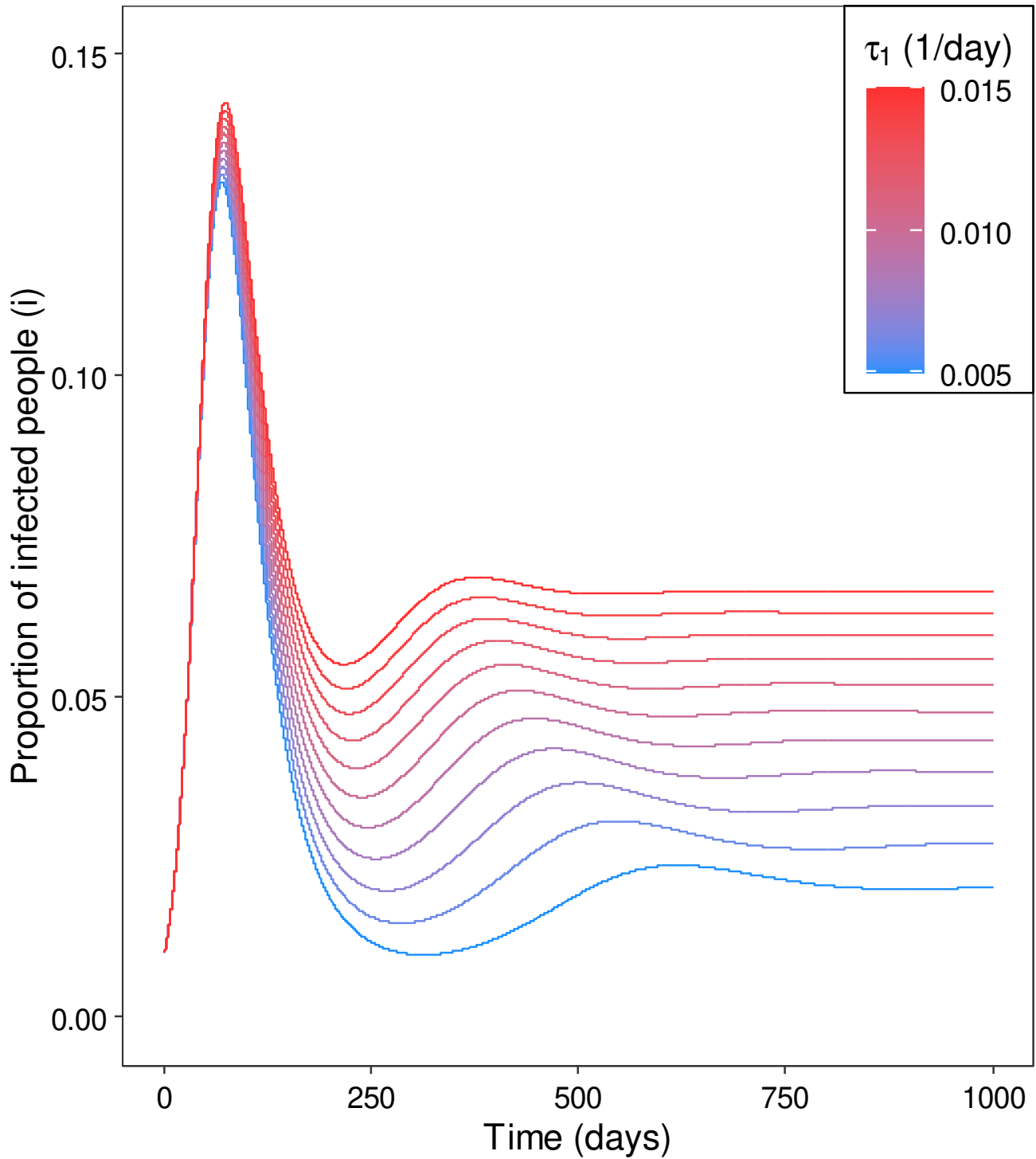

**Figure 10. Influence of the transition rate from $S_L$ to $S_M$ ($\tau_1$) on the evolution of the prevalence ($i$). $\tau_1$ (1/day)={0.005 to 0.015, by increments of 0.001}. The rest of the conditions were maintained as in the benchmark case.**

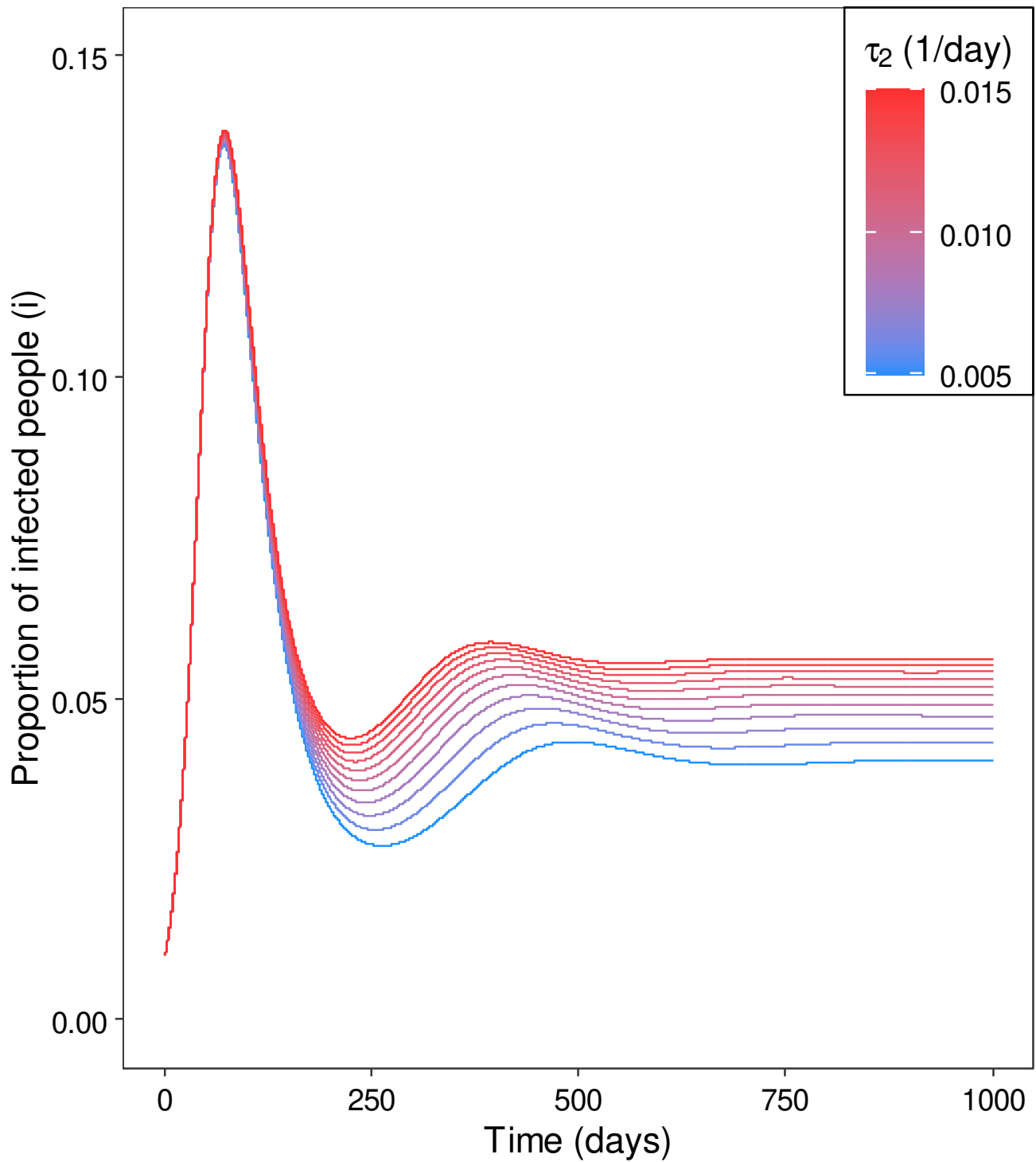

**Figure 11. Influence of the transition rate from $S_M$ to $S_H$ ($\tau_2$) on the evolution of the prevalence ($i$). $\tau_2$ (1/day)={0.005 to 0.015, by increments of 0.001}. The rest of the conditions were maintained as in the benchmark case.**

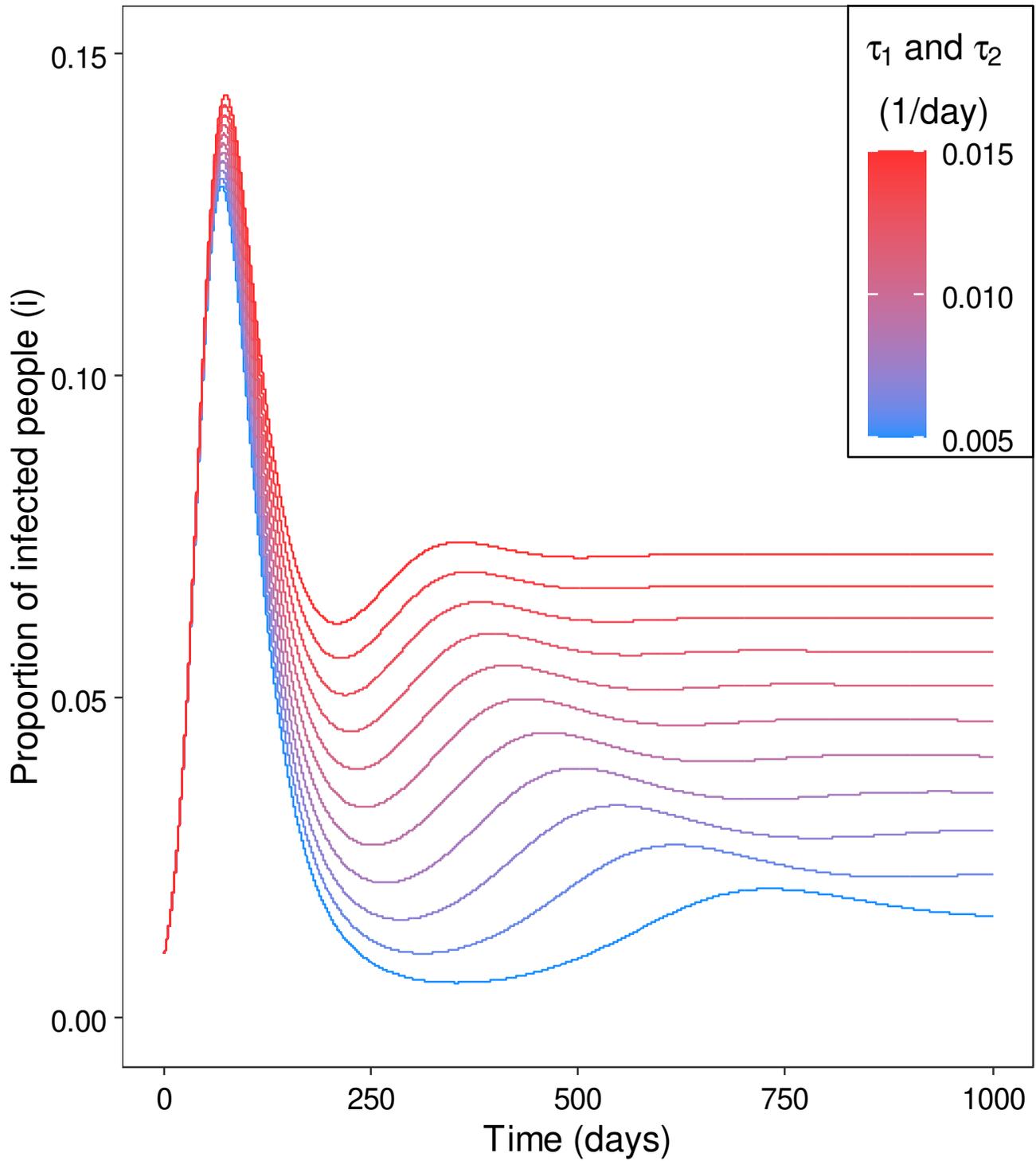

**Figure 12.** Aggregate influence of the transition rates $\tau_1$ ($S_L \rightarrow S_M$) and $\tau_2$ ($S_M \rightarrow S_H$) on the evolution of prevalence (*i*). $\tau_1$ and $\tau_2$ (1/day)={0.005 to 0.015, by increments of 0.001}. The rest of the conditions were maintained as in the benchmark case.

## 4. Discussion

The simulations developed show that the prevalence, after oscillating between peaks and valleys, reaches a plateau phase. In addition, the model shows sensitivity to force of infection, vaccination rate and parameters representing the temporal decay of protection provided by previous infection or vaccination.

The curves of prevalence obtained in the simulations show that the magnitude of the peaks and plateaus increase in different degrees as the infection rate increases, the vaccination rate decreases and the rate of decay of protection conferred by vaccination or previous infection increases.

The magnitude and temporal location of the peaks are of particular epidemiological importance because it could be thought that near the peaks health systems are more compromised [53-55]. For its part, the magnitude of the plateau would provide useful information for dealing with an eventual stable endemic condition [21,56].

The trend towards a certain plateau of prevalence is in line with simulations of other models that have included reinfection or vaccination [57-60].

In order to study the effect of a few variables, we have proposed a very simple model, which could eventually be extended in different ways. Some of the most important limitations of our model are mentioned in the following paragraphs.

Our model proposes three susceptibility stages, which we have named $S_L$, $S_M$ and $S_H$. However, more stages could be incorporated, which could lead to a study on the mathematical properties of a general SISn model, where $n$ would be the number of susceptible stages. Our model would be a particular case with $n=3$.

In the framework of our model, both infected ($I$) and vaccinated susceptible ($S_H$) individuals move into the same compartment ($S_L$). That is, our model assumes that the dynamics of protection are the same for both infected and susceptible vaccinated people. However, there is evidence for differences between the immunity generated by infection and vaccination. While vaccination would produce a higher initial level of antibodies than infection, the decay of antibody levels would be slower with infection [61]. In agreement with this comparison, the results obtained by Gazit *et al.* (2021) suggest that infection would produce greater and longer-lasting protection than vaccination [62]. In this sense, the parameters of our model that relate to the kinetics of protection against infection ($\beta_L$, $\beta_M$, $\beta_H$, $\tau_1$ and $\tau_2$), should be considered as overall measures that summarise the protection generated by previous infection and vaccination.

The model assumes that people achieve the highest protection immediately after vaccination. However, it would take a few days after vaccination to reach the maximum level of immunity [63]. Furthermore, the model does not allow to differentiate between primary vaccination schedules, which may sometimes require more than one dose, and boosters [64,65].

In our hypothetical population, each of the four groups into which the population was divided ($S_L$, $S_M$, $S_H$, $I$) is perfectly homogeneous. However, in the real world, certain variables could cause intragroup heterogeneity. One of these variables would be age. Different studies suggest that the protection conferred by previous infection or vaccination could vary according to age group [8,66,67].

The model does not explicitly represent the emergence of new variants of the virus, which could eventually have higher transmissibility [68]. Thus, the increase in post-infection or post-vaccination susceptibility outlined in the model ($S_L \rightarrow S_M \rightarrow S_H$), could be a consequence of either the natural decay of immunity or the emergence of new variants with higher $R_0$, or a combination of both phenomena. The model neither incorporates non-pharmacological health measures that have been applied almost everywhere in the world, such as isolation, quarantine, social distancing or community containment [69,70].

Our model proposes endless vaccination for the entire population. And, as mentioned above, simulations predict that an increase in the vaccination rate would reduce peak prevalence and produce an endemic phase with fewer infected people. However, beyond this mathematical result presented by our theoretical design, the application of successive boosters is a controversial strategy. One of the main points of debate relates to equity [71-73]. If high vaccination rates in some regions or countries were to substantially reduce access to vaccination in others, a negative effect on global pandemic control could end up occurring. But it should also be borne in mind that not everyone adheres to the vaccine [74-76]. And in particular, some people who received a full primary schedule of vaccination could be resistant to boosters [77].

## 5. Conclusion

The present study suggests that, at least for a certain set of conditions, the spread of SARS-CoV-2, although it could present unstable stages with oscillations, would finally evolve towards an endemic form, with a more or less stable prevalence whose equilibrium value would be regulated by infection and vaccination rates, and by the kinetics of post-infection and post-vaccination protection. However, our development is a theoretical scheme, with simplifications and assumptions that limit its scope. Thus, its predictions should be considered with great care.

**Declaration of competing interest**

The authors declare that they have no conflict of interest.

**Acknowledgments**

The authors gratefully acknowledge Universidad de Buenos Aires for providing financial support (grant number UBACyT 20020190200179BA).